\documentclass[aps,prstab,twocolumn,superscriptaddress,showkey]{revtex4}

\usepackage{graphicx}
\usepackage{amsmath}
\usepackage{dcolumn}
\usepackage{bm}

\makeatletter

\begin{document}

\title{On Retardation Effects in Space Charge Calculations Of High Current Electron Beams}

\author{S.~Becker}
\thanks{Corresponding author}
\email{stefan.becker@physik.uni-muenchen.de}
\author{F.~Gr\"uner}
\author{D.~Habs}
\affiliation{Ludwig-Maximilians-Universit\"at M\"unchen, 85748 Garching, Germany}


\begin{abstract}
Laser-plasma accelerators are expected to deliver electron bunches with high space charge fields. Several recent publications have addressed the impact of space charge effects on such bunches after the extraction into vacuum. Artifacts due to the approximation of retardation effects are addressed, which are typically either neglected or approximated. We discuss a much more appropriate calculation for the case of laser wakefield acceleration with negligible retardation artifacts due to the calculation performed in the mean rest frame. This presented calculation approach also aims at a validation of other simulation approaches.
\end{abstract}

\maketitle

\newcommand{\comment}[1]{}

\section{Introduction}

Laser acceleration is a promising field in various aspects, such as the miniaturization of the accelerator setup and the availability of electron beams with high-current density. The plasma wake field acceleration in the bubble regime was predicted in PIC simulations \cite{bubble} and led to rapid progress in various experiments \cite{nature1,nature2,LeemansNatPhys2,jens}.

The effect of the charge density on the electron beam dynamics after the extraction into vacuum can be estimated when considering the electromagnetic field energy \cite{jackson} per particle in the mean rest frame of the particle bunch
\begin{equation}
u'_{f} = \frac{1}{4\pi\epsilon_{0}N}\sum_{i=2}^{N}\sum_{j=1}^{i-1}\frac{q_{i}q_{j}}{|\vec{r'}_{i}-\vec{r'}_{j}|}
= \frac{\epsilon_{0}}{2N}\int |\vec{E}(\vec{r'})|^{2}d^{3}r',
\label{eqUpot}
\end{equation}
where $N$ is the number of charged particles considered and $q_{i}$ is the charge of the particle $i$ at the position $\vec{r'}_{i}$. Within the mean rest frame, the average electron velocity in the longitudinal direction is at a minimum. For a large N, a homogeneously charged sphere with a radius R scales as $u'_{f} \propto Q^{2}/R$ with Q being the total charge.
Laser accelerated electron beams yield normalized field energies $u_{n}=u'_{f}/m_{e}c^{2}$ which enters a regime which is far above the one that can be reached by conventional accelerators.
Space charge effects in this regime have been examined under various aspects, these are, for example, longitudinal wakefields \cite{bosch,geloni}, energy spreads introduced at the electron extraction from the plasma \cite{strupakov} and the temporal development of the induced energy chirp \cite{sc_gruener}.

A frequently used method to calculate space charge effects consists of simulations based on point-to-point interactions (PPI). Numerical calculations based on PPI are particularly exposed to artifacts if the calculation is performed in the laboratory frame. This issue is discussed in \cite{fubiani} explaining the artifacts in the way PPI simulations commonly account for retardation: In the absence of the knowledge on the particles 4D trajectory, retardation is approximated by assuming constant velocities. Consequently, the acceleration of the simulated particles, i.e.~the macro particles during each finite time step, is neglected. For the regime discussed here, these artifacts are examined in detail. Much more appropriate results can be obtained when the calculation is carried out in the mean rest frame of the electron bunch using PPI. In this case, the average velocities of the macro particles are only weakly relativistic and consequently the retardation artifacts are minimized. This result is then discussed in comparison with other widely applied calculation approaches, which exhibit considerable artifacts which manifest in different characteristics of the longitudinal phase space. Therefore, the design of experiments using laser accelerated electron beams could possibly be misguided, especially table-top Free-Electron Lasers \cite{ttfel} which crucially depend on the characteristics of the longitudinal phase space. The calculations here are performed using GPT \cite{GPT}. The artifacts addressed, however, are not specific to the GPT code, but can be found in any code utilizing the PPI model.

\section{Coulomb Expansion}

The electron bunch considered here is cold and thus has zero divergence and zero emittance. The initial bunch configuration with a spatial Gaussian density distribution in the laboratory frame has the RMS values $\sigma_{x}= \sigma_{y}= \sigma_{z} = 1 \rm{\mu m}$, an initial kinetic energy given by the Lorentz factor $\gamma_{0} = 300$ and a total charge of $Q=1 \rm{nC}$ and hence results in a normalized field energy of $u_{n}\approx 10\%$.

For obtaining a 4D trajectory, we have to make assumptions for the acceleration process: Within the plasma, any expansion driven by the Coulomb forces is suppressed due to the strong plasma fields. We can hence assume that the bunch is in a Gaussian shape before leaving the plasma accelerator in both relevant frames of reference, laboratory frame and the mean rest frame. The end of the plasma is assumed to be a ``sharp'' edge behind which the expanding effect of the Coulomb forces is suddenly switched on. This means that only electrons that have crossed the boundary take effectively part in the Coulomb interaction.

\subsection{Mean Rest Frame}

The mean rest frame for the case considered here is an inertial frame of reference co-propagating with the electron bunch at the constant normalized velocity $\beta_{0}$. Before the Coulomb interaction is switched on, the Lorentz transformation leads to a bunch prolongation of a factor of $\gamma_{0}$. In this frame of reference, the Coulomb interaction cannot set in instantaneously and globally. Instead, the onset of the interaction starts at the front end of the bunch and spreads towards the rear end. Note that the bunch geometry in the mean rest frame leads to a transverse expansion dominating over the longitudinal debunching.

\begin{figure}[ht]
\centerline{
  \centering
   \includegraphics[width=7.8cm]{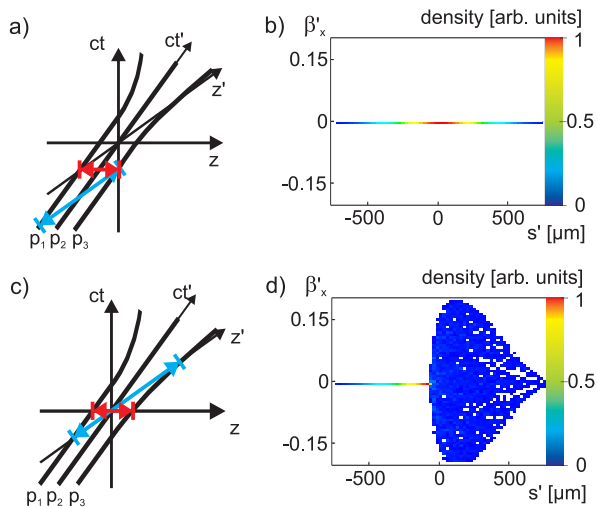}
}
\caption[Bunch When Leaving The Accelerator]{\label{minkowski}The electron bunch is shown schematically in (a) at the end of the wakefield acceleration in its initial configuration using a Minkowski diagram in the laboratory frame (red) and the mean rest frame (blue). The spatial axis z is the longitudinal propagation direction. The plasma boundary is located in the laboratory frame at $z=0$ and in the mean rest frame at $z'=0$. p$_{1}$, p$_{2}$ and p$_{3}$ are 4D trajectories corresponding to the rear end, the center and the front end of the bunch. (b) shows the beam state using a PPI simulation performed in the mean rest frame in correspondence with (a). The longitudinal bunch position $s'$ is plotted against the normalized transverse velocity $\beta'_{x}$ with $s' = 0$ being the center of mass. (c) shows the situation when half of the particles have left the plasma and (d) correspondingly in the mean rest frame at $t' = 0$.}
\end{figure}

Fig.~\ref{minkowski} shows Minkowski diagrams and particle distributions in the mean rest frame of the bunch. Panel \ref{minkowski}a illustrates the initial beam condition in the laboratory frame and in the mean rest frame of the electron bunch before leaving the wakefield accelerator. The bunch is assumed to exhibit a Gaussian spatial density distribution in both frames of reference. This state in the mean rest frame is also shown in panel \ref{minkowski}b using a PPI calculation.

Panel \ref{minkowski}c illustrates the beam in the laboratory frame and in the mean rest frame of the electron bunch at the time $t=t'=0$, where the plasma boundary is at the center of the bunch and moves at the constant velocity $\beta_{\text{pb}}=-\beta_{0}$. This center of the bunch is also determined to be at the longitudinal position zero in both frames $z=z'=0$.

In general we can state that the calculation performed in the laboratory frame does not require sophisticated assumptions concerning the initial simulation conditions, since the predominant transverse space charge interactions can be assumed to set in instantaneously: The duration of propagation of the plasma boundary in the laboratory frame is ultra fast, which means that the shape of the bunch does not significantly change during the propagation of the boundary. Therefore, simulations yield same results for the cases of the electrostatic interaction being switched on instantaneously and the moving of the boundary through the bunch. Within the mean rest frame, however, the different ways of coincidence causes a longitudinal bunch prolongation, which can be seen from panels \ref{minkowski}a,c. Panel \ref{minkowski}d shows the spatial electron density distribution being altered while the plasma boundary is still within the bunch. As a precondition, however, we can assume that the bunch has the same spatial symmetry at the end of the wakefield acceleration in both frames of reference, i.e.~the mean rest frame and the laboratory frame as shown in panel \ref{minkowski}a. Since we have chosen a Gaussian density distribution, we have a spatial point symmetry for the case discussed here. The end of the acceleration distance is at $z = 0$, which requires the introduction of a constraint whereby the onset of the Coulomb interaction propagates from the front end to the rear end of the bunch as shown in panel \ref{minkowski}c.

\begin{figure}[ht]
\centerline{
  \centering
   \includegraphics[width=8.3cm]{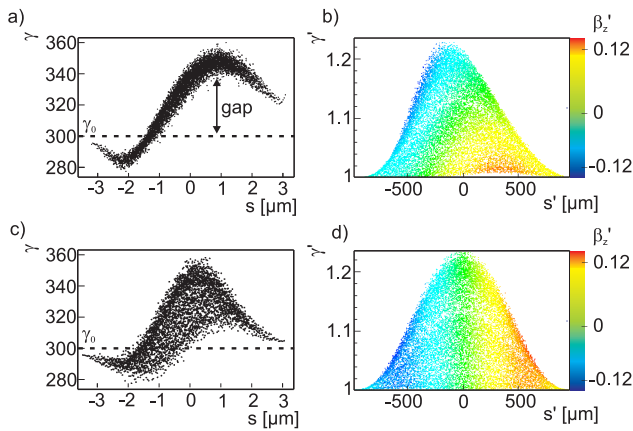}
}
\caption[Rest Frame Asymmetry]{\label{rest}The further evolution of the bunch in Fig.~\ref{minkowski} is shown 3 ps later in the longitudinal phase space with the calculations performed in the mean rest frame. (a) displays the the longitudinal phase space of the bunch transformed to the laboratory frame using Eqs.~\ref{lTrafo}. The slice energy spread yields significantly smaller values than the total energy spread of the bunch. This fact also leads to a gap in the phase space. The reason for this property can be seen in (b), where the longitudinal phase space is drawn with the longitudinal normalized velocity color coded in the mean rest frame. The plasma boundary leads to an asymmetric particle distribution in the mean rest frame. The instantaneous onset of the Coulomb interaction in the mean rest frame would lead to a symmetric particle distribution as shown in (c) and (d) in a direct comparison after 0.7 ps of expansion.}
\end{figure}

\begin{figure*}[ht]
\centerline{
   \includegraphics[width=17cm]{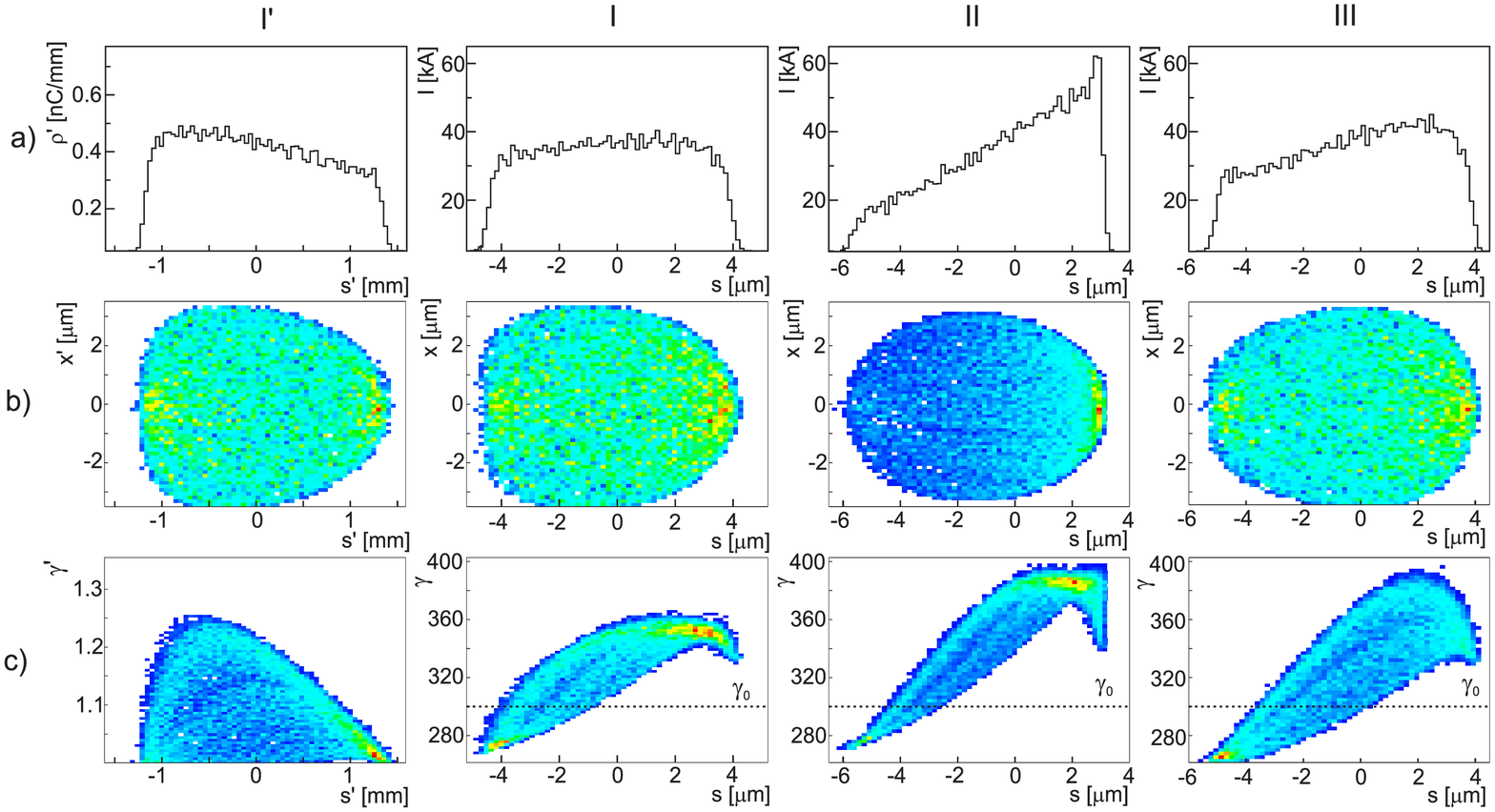}
}
\caption[Spatial density distribution and longitudinal phase space in comparison]{\label{later}Calculation of the vacuum expansion of the considered electron bunch is shown with the initial state as illustrated in Fig.~\ref{minkowski} after the propagation distance of 1.8 m corresponding to 6 ns in the laboratory frame or after 20 ps in the mean rest frame. The center of mass is located at $s=0 = \sum_{i} \gamma_{i}(t)s_{i}(t)/\sum_{i}\gamma_{i}(t)$.
Panels in column I' show the calculation with negligible retardation artifacts in the mean rest frame and panels in column I show the Lorentz-transformed results (Eqs.~\ref{lTrafo}) in the laboratory frame. Column II shows the PPI calculation performed in the laboratory frame and column III shows the calculation result corresponding to a Poisson solver, both with instantaneous onset of the Coulomb interaction. The longitudinal spatial inner bunch position is plotted in the mean rest frame ($s'$) and in the laboratory frame ($s$). Panel I'a shows the line charge density distribution, panels I'b,c show the bunch current. Both of these representations are proportional to the longitudinal particle density. Row (b) shows the spatial particle density distribution with the transverse coordinate plotted in the mean rest frame ($x'$) and in the laboratory frame ($x$). Row (c) shows the longitudinal phase space. The color coding is linear and equivalent to the one of panels (b) and (d) in Fig.~\ref{minkowski}.}
\end{figure*}

The further evolution of the electron bunch calculated in the mean rest frame is shown in the laboratory frame in Fig.~\ref{rest}. The Lorentz transformation into the laboratory frame is performed assuming constant velocity and is given by
\begin{equation}
\begin{array}{rlcrl}
\beta_{xi} =& \frac{\beta'_{xi}}{\gamma_{0}(1+\beta_{0}\beta'_{zi})} &\>& x_{i} =& x'_{i} \\
\beta_{yi} =& \frac{\beta'_{yi}}{\gamma_{0}(1+\beta_{0}\beta'_{zi})} &\>& y_{i} =& y'_{i} \\
\beta_{zi} =& \frac{\beta'_{zi}}{1+\beta_{0}\beta'_{zi}} &\>& z_{i} =&
\frac{z'_{i}}{\gamma_{0}(1+\beta_{0}\beta'_{zi})},
\end{array}
\label{lTrafo}
\end{equation}
where $z_{i}(z'_{i})$ is obtained using linear extrapolation in the $(z,ct)$ space. A slice is a longitudinal subsection of the bunch. The small slice energy spread can be explained regarding the longitudinal phase space in the mean rest frame (panel \ref{rest}b): Particles with highest values of $\gamma'$ originate from the regions with the highest initial electrostatic fields, i.e.~in the longitudinal direction along the axis around the center of the bunch and transversely from off-axis. In contrast to the lab frame, the bunch shape in the mean rest frame changes significantly already during the propagation of the plasma boundary (panel \ref{minkowski}d). The off-axis electrons close to the boundary are predominantly accelerated towards the rear (left) end of the bunch and therefore pushing on-axis electrons towards the head as a result of momentum conservation. Due to the Lorentz transformation $\gamma=\gamma_{0}\gamma'(1+\beta_{z}')$ for a particle at a certain position $s'$, a larger value of $\gamma'$ is reduced by a negative value of $\beta'_{z}$ and a smaller $\gamma'$ is boosted by a positive $\beta'_{z}$. Finally, particles with different values of $\gamma'$ obtain virtually the same $\gamma$ in the laboratory frame, which leads to the ``gap'', i.e.~the small slice energy spread for electrons at a specific bunch position $s$ as can be seen in panel \ref{rest}a.

A symmetric particle distribution in the mean rest frame would be obtained from the instantaneous onset of the Coulomb interaction and is shown in panels \ref{rest}c,d: The phase space does not yield the correlation described above and thus, the slice energy spread is larger.

\section{Comparison Of Calculation Approaches}

The calculation of the space charge driven expansion is examined using PPI according to \cite{jackson,bas,GPT}. The electromagnetic fields are calculated relativistically, where radiation effects are neglected and retardation is treated in accordance with the constant velocity approximation. The Coulomb field of particle $j$ acting on $i$ is given in the rest frame of $j$ by
\begin{align}
\vec{E'}_{j\to i} &= \frac{Q\vec{r'}_{ji}}{4\pi\epsilon_{0}|\vec{r'}_{ji}|^{3}},\\
\vec{r'}_{ji} &= \vec{r}_{ji} + \frac{\gamma_{j}^{2}}{\gamma_{j}+1}(\vec{r}_{ji}\cdot\vec{\beta}_{j})\vec{\beta}_{j} = \vec{r'}_{i} - \vec{r'}_{j},
\end{align}
with $Q$ being the charge of the macro particles and $\vec{r'}_{ji}$ being the distance between the particles in the rest frame of $j$. The Lorentz transformation of the electromagnetic fields of particle $j$ acting on $i$ in the laboratory frame yields

\begin{align}
\vec{E}_{j\to i} =& \gamma_{j}\left[\vec{E'}_{j\to i} -
\frac{\gamma_{j}^{2}}{\gamma_{j}+1}(\vec{\beta}_{j}\cdot\vec{E'}_{j\to i})\vec{\beta}_{j}\right],\nonumber\\
\vec{B}_{j\to i} =& \frac{\gamma_{j}\vec{\beta}_{j}\times\vec{E'}_{j\to i}}{c}.
\label{eqQs}
\end{align}
A tracking code \cite{GPT} applying Eqs.~(\ref{eqQs}) is used to calculate the free drift of the considered electron bunch in vacuum. Fig.~\ref{later} compares results of different calculation methods at a later point in time than Fig.~\ref{rest}. The appropriate calculation using PPI performed in the mean rest frame is shown in the column \ref{later}I'. The results being Lorentz transformed are shown in column \ref{later}I. The slice energy spread (panel \ref{later}Ic) is larger compared to the one shown in panel \ref{rest}a because debunching effects on larger propagation distances cause a longitudinal phase space as displayed in panel \ref{rest}b developing towards the one as in panel \ref{rest}d. Column \ref{later}II shows the PPI calculation performed in the laboratory frame. The difference in comparison with column \ref{later}I can be explained with retardation artifacts due to the constant velocity approximation as described in \cite{fubiani}. A further method besides PPI treating space charge is the application of Poisson solvers, which evaluate the electrostatic space charge field in the mean rest frame of the bunch, where magnetic fields are neglected occurring due to relative velocities. The Lorentz transformation into the laboratory frame introduces the magnetic fields. In this respect, the Poisson solvers referred to solve the equations of motion in the laboratory frame. The calculation time using this method linearly scales with the number of macro particles and thus allows many more macro particles to be considered. We evaluated a calculation using PPI performed in the mean rest frame of the bunch as it would be obtained from a Poisson solver by using the fields
\begin{equation}
\vec{E'}_{j\to i} = \frac{Q\vec{r'}_{ji}}{4\pi\epsilon_{0}|\vec{r'}_{ji}|^{3}},
\>\>\>\>\>\>\>\>\>\>\>
\vec{B'}_{j\to i} = 0.
\label{EPois}
\end{equation}
We obtained virtually identical results for the example bunch considered here comparing the cases utilizing Eqs.~(\ref{eqQs}) and utilizing Eqs.~(\ref{EPois}) in the mean rest frame. Thus, the case examined here yields an appropriate treatment with respect to the retardation artifacts when utilizing a Poisson solver. However, the electromagnetic field being obtained in the mean rest frame also requires the correct consideration of the 4D trajectories of the bunch particles as described above. The result of Poisson solvers assuming instantaneous onset of the Coulomb interaction leads to a longitudinal phase space (panel \ref{rest}c), where the slice energy spread within the laboratory frame is shown to be overestimated compared to the case of correct initial conditions applied (panel \ref{rest}a). To our knowledge, the instantaneous onset of the Coulomb interaction is assumed among the vast majority of codes which are widely applied and which utilize the method of a Poisson solver. In addition, more realistic simulation scenarios might involve particle bunches having a notable energy spread or divergence. In these cases, velocities in the mean rest frame might not allow to neglect the magnetic fields, where PPI codes using Eqs.~(\ref{eqQs}) being performed in the mean rest frame of the bunch yield the least artifacts. Panels \ref{rest}c,d and column \ref{later}III show results as obtained from a Poisson solver at a later point in time. Coulomb interaction is considered in the mean rest frame and is assumed to set in instantaneously. This wrong initial condition leads to the difference compared to column \ref{later}I.
\begin{figure}[ht]
\centerline{
  \centering
   \includegraphics[width=8.5cm]{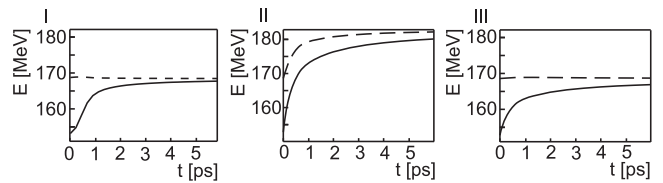}
}
\caption[Total energy]{\label{eTot}The solid line shows the kinetic energy. The dashed line shows the total energy which is the sum of kinetic energy and field energy using Eq.~(\ref{eqUpot}). The calculations correspond to Fig.~\ref{later}.}
\end{figure}
Considering energy conservation for the example beam in Fig.~\ref{eTot} yields the total energy being conserved in the case of the calculation performed in the mean rest frame of the electron bunch and for the case using the Poisson solver.

\section{Conclusion}

The method discussed allows the calculation of the electron bunch evolution in a space charge regime which could be reached by laser acceleration with negligible retardation artifacts. This calculation is performed in the mean rest frame of the bunch, where the relative velocities are only weakly relativistic. Moreover, the required Lorentz transformation between the laboratory frame and the mean rest frame is shown to be non-trivial, since assumptions concerning the 4D trajectories of the particles of the bunch have to be made.
The result of this calculation is compared with the results of two commonly applied methods, one using PPI performed in the laboratory frame and one using a Poisson solver. For both approaches, we found significant deviations concerning the characteristics of the longitudinal phase space, which could ultimately mislead the design of applications using laser accelerated electron bunches:
A PPI simulation performed in the laboratory frame principally suffers from retardation artifacts, which leads to the violation of energy and momentum conservation and to a wrong spatial density distribution. The temporal development of the energy chirp within the longitudinal phase space is overestimated and thus, the examinations in \cite{sc_gruener} describe an upper boundary.
The method of using a Poisson solver in principle suffers from a Lorentz transformation which has to be done in every time step. Some implementations were found to apply the Lorentz transformation for the longitudinal spatial position by merely linearly stretching the bunch by the Lorentz factor corresponding to the velocity of the mean rest frame. Using $z(z')$ as in Eq.~\ref{lTrafo}, instead, helps improving the result. This transformation, however, assumes constant velocity which might introduce artifacts due to the Coulomb driven bunch expansion. The method of using a Poisson solver, hence, could principally deliver better results than PPI in the laboratory frame. However, many simulation codes do not allow to include the correct initial conditions. The method presented here neither requires Lorentz transformations after each time step nor suffers from retardation artifacts. Hence, it offers an appropriate calculation method for the regime of significant space charge effects as expected for laser electron acceleration.

\section*{Acknowledgments}

We thank M. Dohlus (DESY) for fruitful discussions. This work has been funded by the DFG through transregio TR18 and supported by the DFG Cluster-of-Excellence Munich Center for Advanced Photonics MAP.

\label{}

\end{document}